\documentclass{main}
\usepackage{bm,amsmath,amssymb}

\def\be{\begin{equation}}
\def\ee{\end{equation}}
\def\bea{\begin{eqnarray}}
\def\eea{\end{eqnarray}}

\newcommand{\Photo}{}

\begin{document}

\title{\Large  Azimuthal angular correlations in lepton pair production in ultra-peripheral heavy ion collisions
}

\author{Ya-jin Zhou}
\address{Key Laboratory of Particle Physics and Particle Irradiation (MOE), Institute of Frontier and Interdisciplinary Science, Shandong University, QingDao, China}

\maketitle\abstracts{
The coherent photons induced by relativistic heavy ions are highly linearly polarized, in close analogy to the linear polarization of gluons in a large nucleus.  We proposed to measure the photon polarization through azimuthal asymmetries in dilepton production in ultra-peripheral collisions. Our prediction for the asymmetries were soon confirmed by the STAR experiment with high precision. We refined our analysis recently by including the final state soft photon radiation effect beyond the double leading logarithm approximation.  The azimuthal asymmetries and acoplanarity at relatively high transverse momentum provide unique opportunities to test the resummation formalism thanks to the extremely high photon flux in UPCs.  Our results clearly show the feasibility to access the sub-leading resummation effects in UPCs at the RHIC and LHC.
}

\keywords{linearly polarized photon, resummation, ultra-peripheral collisions}

%
%

\section{Linearly polarized gluons and photons}
\begin{figure}[hbpt]
\centering
\includegraphics[scale=0.4]{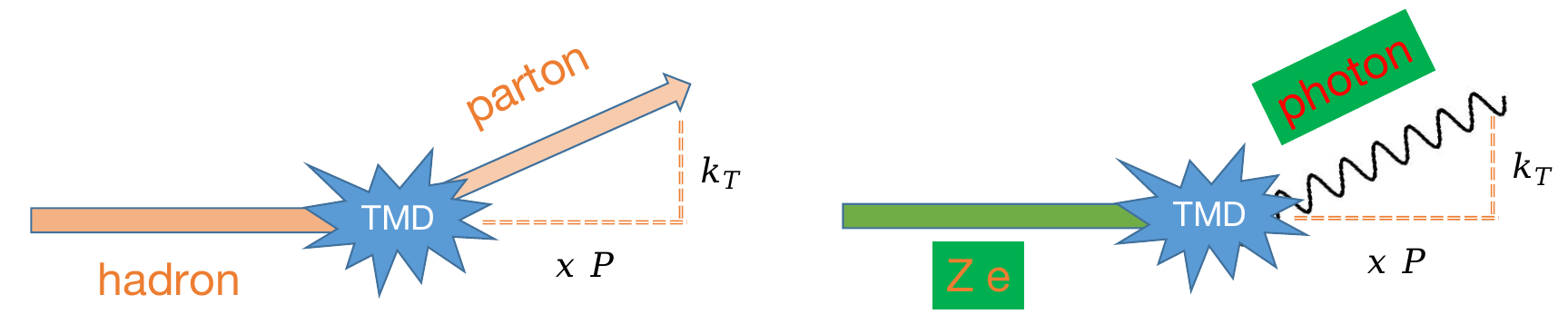}
\caption{Illustration diagrams of the longitudinal and transverse momenta of a parton from a hadron (left) and a photon from a heavy ion (right).} \label{fig:TMDs}
\end{figure}

In high-energy hadron colliders, the partons carry longitudinal momenta as well as transverse momenta, as shown in the left panel of Fig.\ref{fig:TMDs}, which made it necessary to describe the parton distributions in the framework of transverse momentum dependent (TMD) factorization. 
Take the gluon TMD as an example, the leading twist gluon correlator of an unpolarized hadron can be defined and be parametrized in terms of leading twist TMDs
\cite{Mulders:2000sh}:
 \begin{eqnarray} 
 \int \frac{2dy^- d^2y_\perp}{xP^+(2\pi)^3} e^{ik \cdot y} \langle P
|  F_{+}^i(0) U_{[0,y]} F_{+}^j(y) U'_{[y,0]}  |P \rangle \big|_{y^+=0}
=
\delta_\perp^{i j} f_1(x,\bm{k}_\perp^2)+ \left (\frac{2k_\perp^i
k_\perp^j}{\bm{k}_\perp^2}-\delta_\perp^{ij} \right )
 h_1^{\perp }(x,\bm{k}_\perp^2) ,  \label{eq:gTMDs}
\end{eqnarray}
where $f_1(x,\bm{k}_\perp^2)$ and $h_1^\perp(x,\bm{k}_\perp^2)$ are the unpolarized and linearly polarized gluon TMDs, respectively, and the Wilson lines $U_{[0,y]}$ and $U'_{[y,0]}$ guarantee color gauge invariance. For a hadron moves relativistically along $P^+$ direction, $A^+$ component dominant the gauge potential components due to the Lorentz boost effect, so the field tensor reads $F_{+ }^i=\partial_+ A^i-\partial^i A_+$. For a gluon with momentum $k$, the field tensor $F_{+}^i \propto k_{+} A^i-k^i A_{+}$. Usually the transverse momentum of the photon $k_\perp$ is much smaller than its longitudinal momentum $k^+=x P^+$, so it's hard to tell which component is more important in the field tensor. But at small-x limit $k_\perp^i A^+$ is obviously dominant, so $F_{+}^i \propto k_\perp^i A^+ $ approximately. Compare both sides of Eq.\eqref{eq:gTMDs}, one can conclude that $f_1(x,\bm{k}_\perp^2)= h_1^{\perp } (x,\bm{k}_\perp^2)$, which means that the gluons are totally linearly polarized \cite{Metz:2011wb}.

On the other hand, it's well known that relativistically moving ions will introduce electromagnetic field, which can be described by equivalent photon approximation (EPA) developed by Fermi~\cite{Fermi:1924tc}, Weizs\"{a}cker and Williams \cite{vonWeizsacker:1934nji,Williams:1934ad}. These EPA photons will also carry transverse momenta, as shown in the right panel of Fig.\ref{fig:TMDs}, and analogy to gluons, they can also be formulated in the context of TMD factorization using Eq.\eqref{eq:gTMDs}. The only difference is that the Wilson lines guaranteeing gauge invariance in the QCD case are not necessary here, since photons carry no charge and are gauge invariant by themselves. In the typical kinematic region of the heavy-ion colliders, e.g., RHIC and LHC, the EPA photons carry small longitudinal momentum fraction $x$, so based on the analysis in the previous paragraph, it is natural to conclude that the EPA photons are also highly linearly polarized \cite{Li:2019yzy}. 

Linearly polarized gluons can be probed through azimuthal asymmetry effects, so as the linearly polarized photons. Due to the huge QCD background in the central collision area on heavy ion colliders, ultra-peripheral collision (UPC) physics where two heavy ions pass by each other has attracted great interest in recent years. One of the most interesting QED processes in UPCs is $\gamma \gamma \rightarrow l^+ l^-$, which has been extensively studied for unpolarized photons theoretically and experimentally, and is also an ideal process to probe linearly polarized photon through $\cos4\phi$ or $\cos2\phi$ azimuthal asymmetries \cite{Li:2019yzy,Li:2019sin}, as well as a unique way to test the resummation formalism through the all-order resummation of the soft photon radiation effect \cite{Shao:2022stc,Shao:2023zge}.

\section{Probing the linear polarization of photons}
The di-lepton production process via photon-photon fusion at the lowest order QED can be written as
\begin{eqnarray}
\gamma_1(x_1 P+\tilde{k}_{1\perp})+\gamma_2(x_2 \bar{P}+\tilde{k}_{2\perp}) \rightarrow l^+(p_1)+ l^-(p_2),
\end{eqnarray}
where $P$, $\bar{P}$, $p_1$ and $p_2$ represent the four momenta of the two nucleons and the leptons in the final state, respectively. The transverse momenta of the photons are represented by $\bm{k}_{1\perp}$ and $\bm{k}_{2\perp}$, with $\tilde{k}_{1\perp} = (0, 0, \bm{k}_{1\perp})$ and $\tilde{k}_{2\perp} = (0, 0, \bm{k}_{2\perp})$. The leptons are produced in a nearly back-to-back configuration in azimuth with $q_\perp=|\bm{q}_\perp| \ll P_\perp=|\bm{P}_\perp|$, with the total transverse momentum $\bm{q}_\perp\equiv \bm{p}_{1\perp}+\bm{p}_{2\perp}=\bm{k}_{1\perp}+\bm{k}_{2\perp}$ and $\bm{P}_\perp=(\bm{p}_{1\perp}-\bm{p}_{2\perp})/2$.

\begin{figure}[hbpt]
\centering
\includegraphics[scale=0.5]{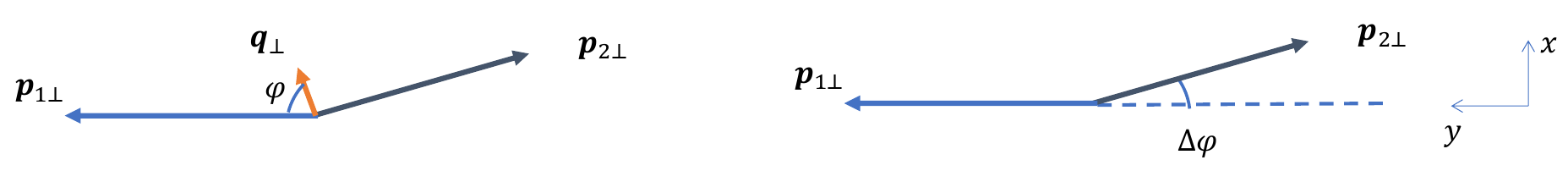}
\caption{Azimuthal angle definitions in the transverse plane.}\label{fig:angle}
\end{figure}

The impact-parameter-integrated cross section is written as
\begin{eqnarray}
\frac{d\sigma}{d^2 \bm{p}_{1\perp} d^2 \bm{p}_{2\perp} dy_1 dy_2}= \frac{2\alpha_e^2}{Q^4}
\left [ \mathcal{A}+ \mathcal{B} \cos 2\phi+\mathcal{C} \cos 4\phi \right ]
\label{eq:cs}
\end{eqnarray}
where $\phi$ is the angle between transverse momenta $\bm{q}_\perp$ and
$\bm{P}_\perp$, as shown in the left panel of Fig.\ref{fig:angle}. 
$y_1$ and $y_2$ are lepton rapidities, respectively. Q is the
invariant mass of the lepton pair.
The coefficients $\mathcal{A}$, $\mathcal{B}$ and $\mathcal{C}$
contain convolutions of photon TMDs, 
 \begin{eqnarray}
\mathcal{A}&=& \frac{(Q^2-2m^2)m^2+(Q^2-2 P_\perp^2)P_\perp^2}{(m^2+P_\perp^2)^2}
x_1x_2\!\int d^2\bm{k}_{1\perp} d^2 \bm{k}_{2\perp} \delta^2( \bm{q_\perp}-\bm{k}_{1\perp}-\bm{k}_{2\perp})
f_1^\gamma(x_1,\bm{k}_{1\perp}^2)f_1^\gamma(x_2,\bm{k}_{2\perp}^2)
\nonumber \\ && +
\frac{m^4}{(m^2+P_\perp^2)^2} x_1x_2 \!
\int \!d^2\bm{k}_{1\perp} d^2 \bm{k}_{2\perp} \delta^2( \bm{q_\perp}\!-\bm{k}_{1\perp}\!-\bm{k}_{2\perp})\!
\left [ 2(\bm{\hat k}_{1\perp} \cdot \bm{\hat k}_{2\perp})^2 \! -1 \right ]
h_1^{\perp \gamma}(x_1,\bm{k}_{1\perp}^2)h_1^{\perp \gamma}(x_2,\bm{k}_{2\perp}^2),  \nonumber \\ ~\\
\mathcal{B}&=&
\frac{4m^2 P_{\perp}^2}{(m^2+P_\perp^2)^2}
x_1x_2\! \int \! d^2\bm{k}_{1\perp} d^2 \bm{k}_{2\perp} \delta^2( \bm{q_\perp}\!-\bm{k}_{1\perp}\!-\bm{k}_{2\perp})
\nonumber \\&& \times  \left \{ \left [ 2(\bm{\hat k}_{2\perp} \! \cdot \bm{\hat q}_{\perp})^2\!-1\right ]
f_1^{ \gamma}\!(x_1,\bm{k}_{1\perp}^2)h_1^{\perp \gamma}\!(x_2,\bm{k}_{2\perp}^2) +
\left [ 2(\bm{\hat k}_{1\perp} \!\cdot \bm{\hat q}_{\perp})^2\!-1\right ]
h_1^{\perp \gamma}\!(x_1,\bm{k}_{1\perp}^2)f_1^{\gamma}\!(x_2,\bm{k}_{2\perp}^2)
\right \}, \\
\mathcal{C} &=&
\frac{ -2P_{\perp}^4}{(m^2+P_\perp^2)^2}
x_1x_2\! \int \! d^2\bm{k}_{1\perp} d^2 \bm{k}_{2\perp} \delta^2( \bm{q_\perp}\!-\bm{k}_{1\perp}\!-\bm{k}_{2\perp})
\nonumber \\
&&\times \left [2\left (\! 2(\bm{\hat k}_{2\perp} \! \cdot \bm{\hat q}_{\perp})(\bm{\hat k}_{1\perp} \! \cdot \bm{\hat q}_{\perp})
-\bm{\hat k}_{1\perp} \!\cdot \!\bm{\hat k}_{2\perp} \! \right )^2\!-1\right ]
 h_1^{ \perp \gamma}\!(x_1,\bm{k}_{1\perp}^2) h_1^{\perp \gamma}\!(x_2,\bm{k}_{2\perp}^2),
\end{eqnarray}
where a vector with a hat on it means a unit vector, and $f^\gamma_1(x,\bm{k}_\perp^2)$ and $h_1^{\perp\gamma}(x,\bm{k}_\perp^2)$ represent the unpolarized and linearly polarized photon TMDs, respectively. One can clearly see that the process has unique angular correlations induced by linearly polarized photons. The longitudinal momentum fractions of the leptons are fixed according to $x_{1} \simeq\sqrt{(P_{\perp}^{2}+m^{2})/s} \left(e^{y_{1}}+e^{y_{2}}\right)$,
$x_{2} \simeq\sqrt{(P_{\perp}^{2}+m^{2})/s}\left(e^{-y_{1}}+e^{-y_{2}}\right)$, with $s$, $m$ being the center of mass energy and the lepton mass, respectively.

To sort out the UPC events, we must include the dependence of the impact parameters in the cross-section, and then integrate $\bm{b}_\perp$  from $2R_{\mathrm{WS}}$ to $\infty$, where $\bm{b}_\perp$ is the impact parameter between the two colliding nuclei and $R_{\mathrm{WS}}$ is the nuclear radius. Once $\bm{b}_\perp$ is introduced, the incident coherent photon is no longer in the eigenstate of the transverse momentum, and accordingly the photon transverse momenta appearing in the amplitude and conjugate amplitude are no longer the same. We use $\bm{k}_{1\perp}$, $\bm{k}_{2\perp}$ and $\bm{k}_{1\perp}'$, $\bm{k}_{2\perp}'$ to denote the transverse momenta in the amplitude and in the conjugate amplitude with the constraint
 $ \bm{k}_{1\perp}'+\bm{k}_{2\perp}'\equiv \bm{q}_\perp$. The calculation of the impact parameter dependent cross-section was first developed in Ref. \citelow{Vidovic:1992ik}, and we extended this calculation to the azimuthally dependent situation \cite{Li:2019sin}. The joint $\bm{b}_\perp$ and $\bm{q}_\perp$ dependent di-lepton production cross section at the lowest order of QED can be written as,
\begin{eqnarray}
 \frac{d \sigma_{0}}{d^{2} \bm{p}_{1\perp} \bm{p}_{2\perp} d y_{1} d y_{2} d^{2} \bm{b}_{\perp}} = \frac{2\alpha_e^2}{(2\pi)^2Q^4}
\left [ \mathcal{A}+ \mathcal{B} \cos 2\phi+\mathcal{C} \cos 4\phi \right ]
\end{eqnarray}\label{born}
with
\begin{eqnarray}
\mathcal{A}\! & = & \!{\cal \int}[{d\cal K}_\perp]\frac{1}{\left(P_{\perp}^{2}+m^{2}\right)^{2}} \Bigl[-2m^{4}\cos\left(\phi_{\bm{k}_{1\perp}}+\phi_{\bm{k}_{1\perp}'}-\phi_{\bm{k}_{2\perp}}-\phi_{\bm{k}_{2\perp}'}\right)+m^{2}\left(Q^{2}-2m^{2}\right) \nonumber \\ &&\times 
\cos\left(\phi_{\bm{k}_{1\perp}}-\phi_{\bm{k}_{1\perp}'}-\phi_{\bm{k}_{2\perp}}+\phi_{\bm{k}_{2\perp}'}\right)
 +P_{\perp}^{2}\left(Q^{2}-2P_{\perp}^{2}\right)\cos\left(\phi_{\bm{k}_{1\perp}}-\phi_{\bm{k}_{1\perp}'}+\phi_{\bm{k}_{2\perp}}-\phi_{\bm{k}_{2\perp}'}\right)\Bigr],  \\
\mathcal{B} \! & = & \!{\cal \int}[{d\cal K}_\perp]\frac{8m^{2}P_{\perp}^{2}}{ \left(P_{\perp}^{2}+m^{2}\right)^{2}} \cos\left(\phi_{\bm{k}_{1\perp}}-\phi_{\bm{k}_{2\perp}}\right)\cos\left(\phi_{\bm{k}_{1\perp}'}+\phi_{\bm{k}_{2\perp}'}-2\phi\right),\\
\mathcal{C} \! & = & \!{\cal \int}[{d\cal K}_\perp]\frac{-2P_{\perp}^{4}}{\left(P_{\perp}^{2}+m^{2}\right)^{2}}\cos\left(\phi_{\bm{k}_{1\perp}}+\phi_{\bm{k}_{1\perp}'}+\phi_{\bm{k}_{2\perp}}+\phi_{\bm{k}_{2\perp}'}-4\phi\right), 
\end{eqnarray}
where $\phi_{\bm{k}_{i}}$ is the azimuthal angle between $\bm{P}_\perp$ and $\bm{k}_{i}$, and the shorthand notation represent
\begin{eqnarray}\nonumber
{\cal \int}[d{\cal K}_\perp ]&\equiv& \int d^{2}\bm{k}_{1\perp}d^{2}\bm{k}_{2\perp}d^{2}\bm{k}_{1\perp}'d^{2}\bm{k}_{2\perp}'e^{i(\bm{k}_{1\perp}-\bm{k}_{1\perp}')\cdot \bm{b}_{\perp}} 
\delta^{2}(\bm{k}_{1\perp}+\bm{k}_{2\perp}-\bm{q}_{\perp}) \delta^{2}(\bm{k}_{1\perp}'+\bm{k}_{2\perp}'-\bm{q}_{\perp}) \\ && \times 
\mathcal{F}(x_1,\bm{k}_{1\perp}^{2})\mathcal{F}(x_2, \bm{k}_{2\perp}^{2})\mathcal{F}(x_1, \bm{k}_{1\perp}'^{2})\mathcal{F}(x_2, \bm{k}_{2\perp}'^{2}). 
\end{eqnarray}
One notices that the $\bm{b}_\perp$ dependence enters the cross section via the phase $e^{i(\bm{k}_{1\perp}-\bm{k}_{1\perp}')\cdot \bm{b}_{\perp}} $. The function $\mathcal{F}(x_{1}, \bm{k}_{\perp}^{2})$ describes the probability amplitude for a photon carrying a given momentum. It can be related to the normal photon TMD:  
\begin{eqnarray}
    |\mathcal{F}(x_{1}, \bm{k}_{\perp}^{2})|^2=x_1 f_1^\gamma(x_1,\bm{k}_{\perp}^2)=
\frac{Z^2 \alpha_e}{\pi^2} \bm{k}_{\perp}^2
\left [ \frac{F(\bm{k}_{\perp}^2+x^2M_p^2)}{(\bm{k}_{\perp}^2+x^2M_p^2)}\right ]^2,
\label{f1h1}
\end{eqnarray}
where $F$ is the Woods-Saxon form factor
\begin{eqnarray} \
F(\bm k^2)= \int d^3 \bm{r} e^{i\bm k\cdot \bm r} \frac{\rho^0}{1+\exp{\left [(r-R_{\mathrm{WS}})/d\right ]}}.
\end{eqnarray}

Considering higher-order QED contributions, the lepton pair will acquire a recoil transverse momentum through the final state soft photon radiation effect, which will cause large logarithmic terms $\alpha_e^n {\rm ln}^{2n} \frac{Q^{2}}{m^2} $. These large logarithms can be resumed by using the Collins-Soper formalism \cite{Collins:1984kg} and result in Sudakov factor in the exponential in the impact parameter space. At the one-loop order the Sudakov factor reads \cite{Klein:2018fmp,Hatta:2021jcd},
\begin{eqnarray}\label{eq:sud_DL} 
  \mathrm{Sud}(\mu_r,r_\perp)=
\frac{\alpha_e}{\pi} {\rm ln} \frac{Q^2}{m^2}  {\rm ln}\frac{P_\perp^2}{\mu_r^2},
\end{eqnarray}
with $\mu_r=2 e^{-\gamma_E}/r_{\perp}$. The cross section is then expressed as
\begin{eqnarray}
  \frac{d\sigma}{d^2 \bm{p}_{1\perp} d^2 \bm{p}_{2\perp} dy_1 dy_2 d^2 \bm{b}_\perp }= \int
  \frac{d^2 \bm{r}_\perp}{(2\pi)^2} e^{i \bm{r}_\perp \cdot \bm{q}_\perp} e^{- \mathrm{Sud}(r_\perp)} \int d^2 \bm{q}_\perp'
  e^{-i \bm{r}_\perp \cdot \bm{q}_\perp'} \frac{d\sigma_{_{\!0}}(\bm q_\perp')}{ d\mathcal{P.S.}}, \label{eq:res1}
\end{eqnarray}
where $d {\cal P.S.}=d^2 \bm{p}_{1\perp} d^2 \bm{p}_{2\perp} dy_1 dy_2 d^2 \bm{b}_\perp $ being the phase space factor. 
The azimuthal asymmetries, i.e., the average value of the $\cos(n\phi)$, is defined as
\begin{eqnarray}
\langle \cos(n\phi) \rangle =\frac{ \int \frac{d \sigma}{d {\cal P.S.}} \cos (n\phi) \ d {\cal P.S.} }
{\int \frac{d \sigma}{d {\cal P.S.}}  d {\cal P.S.}}.
\end{eqnarray}

\begin{figure}[hpbt]
    \centering
    \includegraphics[scale=0.5]{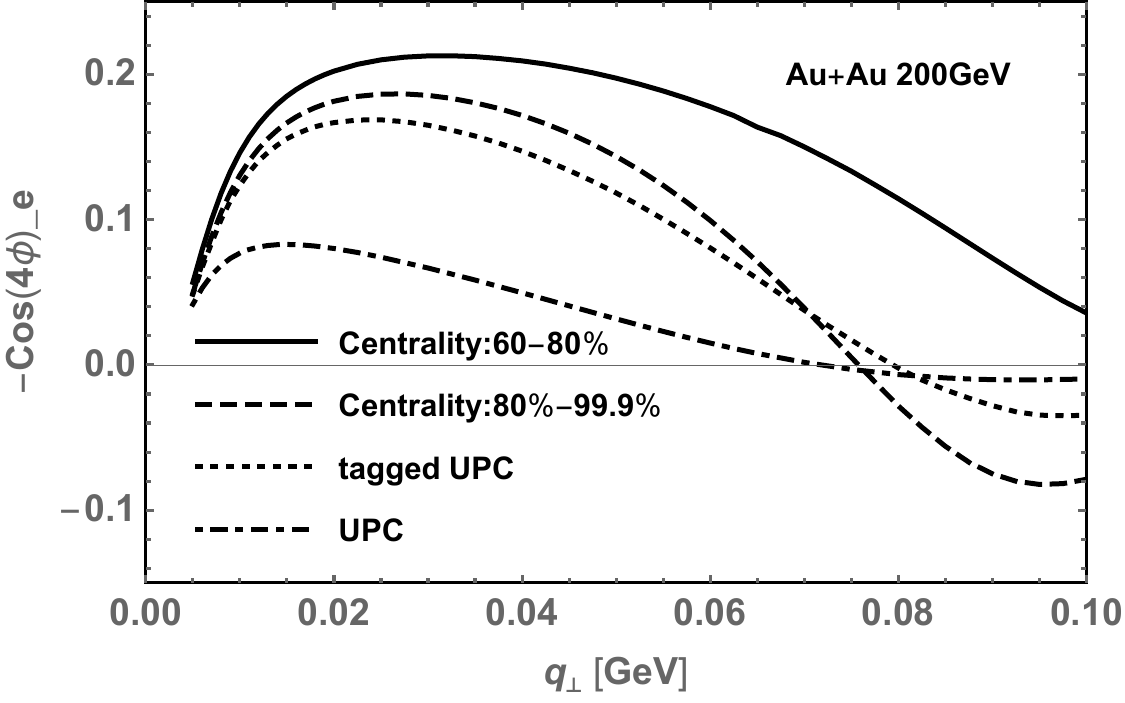}
    \caption{ Estimates of the $\cos 4\phi$ asymmetry as the function of $q_\perp$ for different centralities at $\sqrt{s}=200~\mathrm{GeV}$. The electron and positron rapidities and transverse momenta are integrated over the regions [-1, 1] and [0.2, 0.4] GeV, respectively.}
 \label{fig:C4Au}
\end{figure}

Numerical results for $\langle \cos(4\phi) \rangle$ of the di-electron and di-muon production in UPCs at RHIC and LHC energy regions have been calculated in Refs. \citelow{Li:2019yzy} and \citelow{Li:2019sin}, where we predicted significant $\langle \cos(4\phi) \rangle$ asymmetries as shown in Fig.\ref{fig:C4Au}.
This effect was promotely verified by RHIC STAR collaboration \cite{Adam:2019mby}. As demonstrated in Table \ref{tab:v4}, the measured $\langle \cos(4\phi) \rangle$ for the process $\gamma \gamma \to e^+e^-$ aligns excellently with the theoretical predictions. A similar theoretical outcome for the centrality of 60\%-80\% was obtained based on the Wigner distributions of photons \cite{Klusek-Gawenda:2020eja}.
\begin{table}[hbpt]
    \centering
    \begin{tabular}{|c|c|c|}
    \hline
         & Measured $|2\langle \cos(4\phi) \rangle|$ & QED calculated $2\langle \cos(4\phi) \rangle$\\
    \hline
     UPC & 16.8\%$\pm$2.5\% & $-$16.5\% \\
    \hline
60\%-80\%& 27\%$\pm$6\%     & $-$34.5\%  \\
    \hline
    \end{tabular}
    \caption{Theoretical and experimental results for $\cos(4\phi)$ asymmetry in di-electron UPC production in Au+Au collisions at RHIC with $\sqrt{s} = 200 ~\mathrm{GeV}$. The invariant mass Q is integrated over [0.45, 0.76] GeV, the electron and positron rapidities $y_{1,2}$ are integrated over [-1, 1], and the total transverse momentum of the dielectron $q_\perp$ is integrated over [0, 0.1] GeV. The individual electron transverse momentum is restricted to $P_\perp > 0.2 ~\mathrm{GeV}$.}
    \label{tab:v4}
\end{table}

\section{Toward the precision test of the resummation formalism}
Besides directly measuring the $q_\perp$ distribution, the azimuthal angular decorrelation of the lepton pair is often experimentally studied as well. When the lepton pair acquires finite transverse momentum, they will deviate from the exact back-to-back configuration in the transverse momentum space, and the deviation degree is measured by the so called acoplanarity with the definition being $\alpha= |\Delta\phi|/\pi $. The azimuthal angle $\Delta\phi$ is defined as $\Delta\phi =\pi -( \phi_1 - \phi_2)$ where $\phi_1$ and $\phi_2$ represent the azimuthal angles for the lepton and the anti-lepton, respectively, as shown in the right panel of Fig.\ref{fig:angle}. We fix the direction of the electron transverse momentum $\bm{p}_{1\perp} $ to be Y-axis.  The acoplanarity can then be easily reconstructed by the ratio of $q_x$ (the component of $\bm{q}_\perp$ aligned with X-axis) and $P_\perp$.

In this section we discuss the soft photon radiation effect on both the azimuthal asymmetries and the acoplanarity. 
In the previous studies the soft photon contributions were resummed within double leading logarithm approximation \cite{Klein:2018fmp,Klein:2020jom}.
In Ref. \citelow{Shao:2023zge}, we extended the resummation formalism to the next to leading logarithm accuracy and investigate its phenomenological consequence as well. We will briefly demonstrate the steps and results later. 

\subsection{The calculation of the azimuthal asymmetries}
First we discuss the azimuthal asymmetry.
It's already known that the final state soft photon radiation effect introduces modifications to the cross sections, as discussed in the preceding section, thereby influencing the magnitude of azimuthal asymmetries. While on the other hand, the photon that is emitted also induce a recoil effect on the lepton that emit it. Given that the lepton pair are nearly back-to-back in the transverse plane, the recoil effect will also cause azimuthal asymmetry \cite{Hatta:2020bgy,Hatta:2021jcd}. The cross section with soft photon radiation can be written as \cite{Catani:2014qha,Catani:2017tuc,Hatta:2020bgy,Hatta:2021jcd}
\begin{eqnarray}
    \frac{d \sigma(q_\perp)}{d {\cal P.S.}}=\int d^2 q_\perp' \frac{d \sigma_0(q_\perp')}{d {\cal P.S.}}
    S(q_\perp-q_\perp')
\end{eqnarray}
and the soft factor is expanded at the leading order as~\cite{Hatta:2021jcd}, 
\begin{eqnarray} 
S(l_{ \perp})\!=\! \delta(l_{ \perp})+ \frac{\alpha_e } {\pi^2 l_{ \perp}^2} \left \{ c_0+2\,c_2 \cos 2\phi+2\,c_4 \cos 4\phi+... \right \}, 
\label{inte}
\end{eqnarray}
with $c_0\approx \ln \frac{Q^2}{m_\pi^2}$, $c_2\approx \ln \frac{Q^2}{m_\pi^2}+\delta y \sinh \delta y-2\cosh^2 \frac{\delta y}{2} \ln [2(1+\cosh \delta y)]$... when the final-state particle mass is
much smaller than $P_\perp$. $\delta y=y_1-y_2$ is the difference between the two rapidities of the leptons.

The final state soft photon radiation mainly occurs in regions where the lepton transverse momentum $q_\perp$ is relatively large, for instance, when $q_\perp$ is larger than about 100 MeV. The previous section mainly focused on the small transverse momentum area, so we ignore the azimuthal asymmetry effect induced by the recoiled lepton there. In this section we will cover the [0, 200] MeV region for $q_\perp$, where both coherent photons and final state radiations can play their important roles. Considering an all-order resummation of the large logarithmic terms, the cross section is also expressed as Eq.\eqref{eq:res1} in the transverse position space, except that here we will include the subleading logarithm contributions.

 We used the soft collinear effective theory (SCET) \cite{Bauer:2000yr,Bauer:2001ct,Bauer:2001yt,Bauer:2002nz,Beneke:2002ph} and the standard Renormalization Group (RG) methods to derive the resummation formula that includes the effects of lepton mass resummation to all orders. We re-factorize the massive hard and soft functions in the small mass limit ($Q \gg q_\perp \gtrsim m$), where the massive hard function $H(Q,m,\mu)$ is factorized as the product of the massless hard function $H(Q,\mu)$ and collinear jet functions $J(m,\mu)$, and the massive soft function $S(l_\perp,\Delta y,m,\mu)$ is factorized as the product of the massless soft function $S(l_\perp,\Delta y,\mu)$ and collinear-soft functions $C_{i}(k_{i,\perp},p_T,m,\mu)$. 
 The resulting differential cross section is given by
\begin{align}\label{eq:fac_qT}
	\frac{d\sigma(q_\perp)}{ d\mathcal{P.S.}} = & \, H(Q,\mu) J^2(m,\mu) \int d^2 \bm{l}_\perp d^2 \bm{k}_{1\perp} d^2 \bm{k}_{2\perp}  \frac{d\sigma_0(\bm{q}_\perp-\bm{l}_\perp - \bm{k}_{1\perp }- \bm{k}_{2\perp} )}{ d\mathcal{P.S.}}  \notag \\
&\times S(l_\perp,\Delta y,\mu) C_1 ( k_{1\perp },P_\perp,y_1,m,\mu ) C_2 ( k_{2\perp },P_\perp,y_2,m,\mu ),  
\end{align}
where the hard function $H(Q,\mu)$ comes from the matching from QED to the low energy effective theory, and the corresponding anomalous dimension is written as
\begin{align}
    \Gamma_H &= \frac{\alpha_e}{4\pi} \left( 8 \ln \frac{Q^2}{\mu^2} - 12\right).
\end{align}
One can calculate the next-to-leading order contributions to the collinear jet function, the massive soft function, and the collinear-soft function, respectively, and obtain the corresponding anomalous dimensions which satisfying the consistency relations for the RG evolutions. The results are:
\begin{align} 
      \Gamma_S &= \frac{\alpha_e}{4\pi}\left( 8 \ln \frac{\mu^2r_\perp^2}{b_0^2} + 8 \ln \cos^2\phi_r - 8 \ln\frac{1+\cosh\Delta y}{2}\right), \\ 
 \Gamma_J &= \frac{\alpha_e}{4\pi} \left( 4 \ln \frac{\mu^2}{m^2} + 2\right),\\
        \Gamma_{C_{1,2}} &= \frac{\alpha_e}{4\pi}\left( - 4 \ln \frac{4P_\perp^2\mu^2r_\perp^2}{b_0^2 m^2} + 4 - 4 \ln \cos^2\phi_r \pm 4 i \pi  \right).  
\end{align}

Then the Sudakov factor is given by
\begin{align}
	\mathrm{Sud}(r_\perp) = \int_{\mu_r}^Q \frac{d\mu}{\mu} \Gamma_H + 2\int_{\mu_r}^m \frac{d\mu}{\mu} \Gamma_J + \int_{\mu_r}^{\mu_r m/(2P_\perp)} \frac{d\mu}{\mu} \Gamma_{C_{1}} + \int_{\mu_r}^{\mu_r m/(2P_\perp)} \frac{d\mu}{\mu} \Gamma_{C_{2}}. 
 \label{Res_all}
\end{align}

\begin{figure}[htpb]\centering
    \includegraphics[scale=0.4]{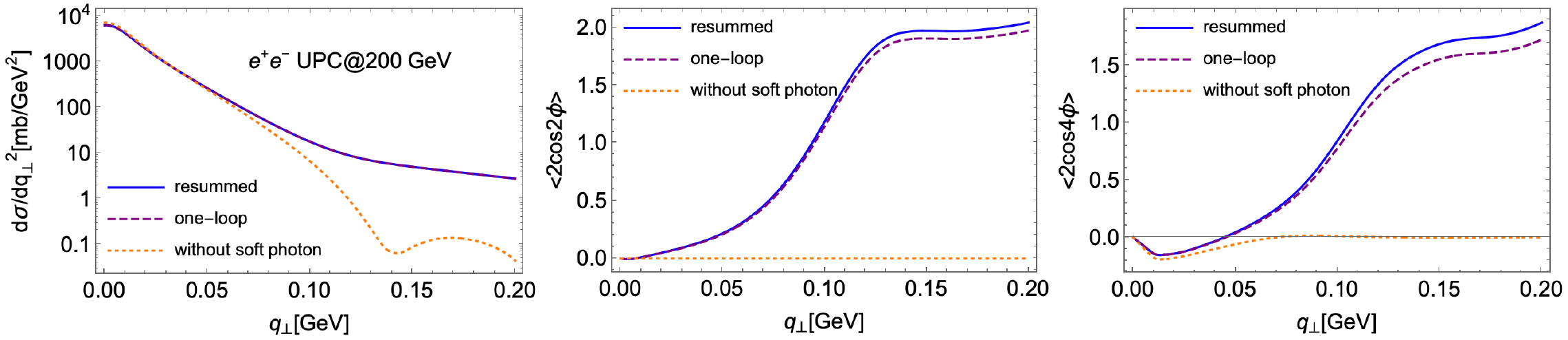}
    \caption{Di-electron production in unrestricted UPCs in Au+Au collisions at the RHIC energy. The following kinematic cuts are imposed: the rapidities $|y_{1,2}|<1$, transverse momentum $P_\perp >200$ MeV, and the invariant mass of the electron pair $450~\text{MeV}< Q < 760~\text{MeV}$. Left panel: azimuthal averaged differential cross sections; middle panel: $\langle\cos(2\phi)\rangle$ azimuthal asymmetry; right panel: $\langle\cos(4\phi)\rangle$ azimuthal asymmetry.}
    \label{fig:aziSTAR}
\end{figure}

\begin{figure}[htpb]\centering
    \includegraphics[scale=0.4]{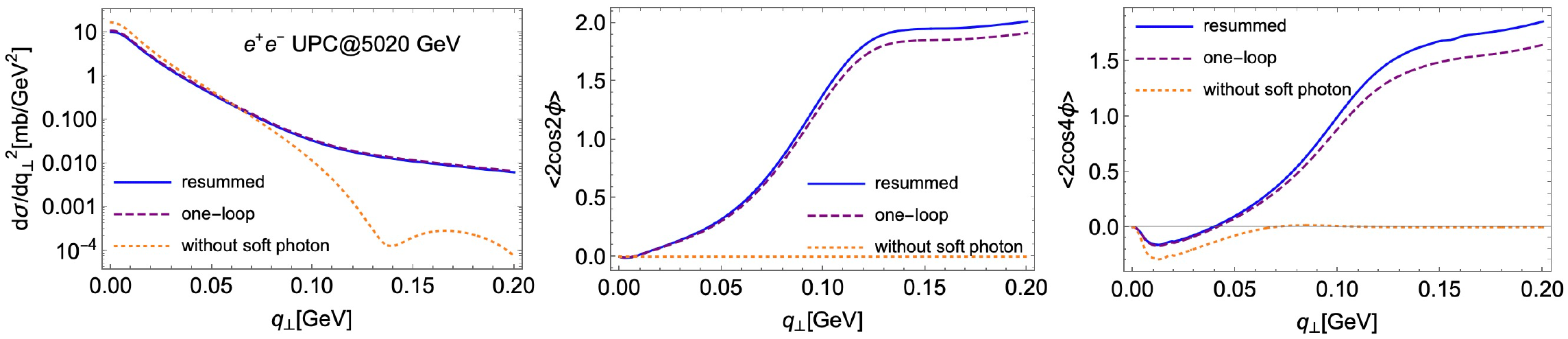}
    \caption{Di-electron production in unrestricted UPCs in Pb+Pb collisions at the LHC energy. The following kinematic cuts are imposed: the rapidities $|y_{1,2}|<0.8$ and the invariant mass of the di-electron $10~\text{GeV}< Q < 20~\text{GeV}$. Left panel: azimuthal averaged differential cross sections; middle panel: $\langle\cos(2\phi)\rangle$ azimuthal asymmetry; right panel: $\langle\cos(4\phi)\rangle$ azimuthal asymmetry.}
    \label{fig:aziATLAS}
\end{figure}

In the numerical calculation we compute both the azimuthal independent cross sections and the $\cos2\phi$ and $\cos4\phi$ asymmetries for the unrestricted UPC case, where the impact parameter is simply integrated over 
$[2R_{\mathrm{WS}}, \infty)$. The nucleus radius $R_{\mathrm{WS}}$ is taken as 6.4 fm for Au and 6.68 fm for Pb. We show the azimuthal independent cross section, $\langle (\cos 2\phi)\rangle$ and  $\langle (\cos 4\phi)\rangle$ as a function of $q_\perp$ at RHIC energy in Fig.\ref{fig:aziSTAR} and at LHC energy in Fig.\ref{fig:aziATLAS}. In these curves the blue solid lines stand for the fully resummed results from Eq.\eqref{Res_all}, and the purple dashed lines represent the results with the azimuthal dependent part being treated at the one-loop order. The results without soft photon radiation effect are shown with the dotted orange lines. It is evident that the perturbative tail, which is a result of soft photon radiation, takes precedence over the lepton pair transverse momentum spectrum, which is determined by the coherent photon primordial $k_\perp$ distribution, at relatively high $q_\perp$ values. The most remarkable contrast is the $\cos2\phi$ asymmetry, which is ignorable at the lowest order due to the negligible mass of electron, exhibit large values at high $q_\perp$. In the current study, we resum both the azimuthally independent and dependent leading logarithms into an exponential form. In the previous works~\cite{Hatta:2020bgy,Hatta:2021jcd}, the azimuthally independent logarithm was resummed to all orders and the azimuthally dependent component was treated at the fixed order. We make a numerical comparison of the results derived from these two resummation schemes. The difference between these two approaches becomes apparent in the large $q_\perp$ region, especially for the $\cos 4\phi$ azimuthal asymmetry. It would be intriguing to examine such a resummation effect in future experiments. 

\subsection{The calculation of the acoplanarity}
In the coordinate frame that we established at the beginning of this section, the acoplanarity can be constructed as $\alpha=q_x/P_\perp$. Integrate Eq.\eqref{eq:res1} over $q_y$ one obtain the $q_x$ dependent cross section,
\begin{eqnarray}
  \frac{d\sigma}{dq_x d^2 \bm{P}_{\perp} dy_1 dy_2 d^2 \bm{b}_\perp }&=& \int dq_y
  \frac{dr_y d r_x}{(2\pi)^2} e^{i( r_x  q_x+r_y  q_y)} e^{-  \mathrm{Sud_a}(r_x,r_y)} \int d q_x' dq_y' \ 
  e^{-i (r_x  q_x'+r_y  q_y')} \frac{d\sigma_{_{\!0}}(q_\perp')}{ d\mathcal{P.S.}} \nonumber \\
  &=& \int 
  \frac{d r_x}{2\pi} e^{i r_x  q_x} e^{-  \mathrm{Sud_a}(r_x,r_y=0)} \int d q_x' dq_y' \ 
  e^{-i r_x  q_x'} \frac{d\sigma_{_{\!0}}(q_\perp')}{ d\mathcal{P.S.}},\label{eq:resa}
  \end{eqnarray}
  where the leading logarithm contribution to the Sudakov factor $ \mathrm{Sud_a}(r_x)$ is given by,
\begin{align}\label{eq:suda_DL}
 \mathrm{Sud_a}(r_x) = \frac{\alpha_e}{2\pi} \left[ \ln^2 \frac{Q^2}{\mu_{rx}^2}  - \ln^2 \frac{m^2}{\mu_{rx}^2}  \theta(m-\mu_{rx})\right] ,
\end{align}
 with $\mu_{rx}=2 e^{-\gamma_E}/r_x$.

A one-dimensional Fourier transform is made in Eq.\eqref{eq:resa} since the acoplanarity is a one-dimensional observable, apart from the two-dimensional $q_\perp$ distribution situation. When deriving the momentum space expression of the Sudakov factor $\mathrm{Sud_a}(l_x)$, the Y-component of the soft photon transverse momentum has to be integrated over the whole available phase-space region. Given this we need to re-derive a factorization formula instead of integrating out $q_y$ from the resummed $q_\perp$ distribution. At the small $\alpha$ limit, the factorization formula is written as,
\begin{align}\label{eq:fac_qx}
	\frac{d\sigma(\alpha)}{ d\mathcal{P.S.}} =& 2 P_\perp H(Q,\mu) J^2(m,\mu) \int d l_x d k_{1,x} d k_{2,x}  \frac{d\sigma_0(q_x-l_x - k_{1x }- k_{2x} )}{ d\mathcal{P.S.}} \notag \\
 &\times S(l_x,\Delta y,\mu,\nu)  C_1 ( k_{1x },P_\perp,y_1,m,\mu,\nu ) C_2 ( k_{2x },P_\perp,y_2,m,\mu,\nu ),  
\end{align}
where the hard and jet functions are the same with those in Eq.\eqref{eq:fac_qT} since they are independent of specific observables. Implementing the one-dimensional Fourier transformation, one can define the $r_x$ dependent soft and soft-collinear functions, calculate them to the next-to-leading order, and finally resummed to all orders as exponential form. The Sudakov factor is expressed as
\begin{align}\notag 
    \mathrm{Sud}_a(r_x) & = \int_{\mu_{rx}}^Q \frac{d\mu}{\mu} \Gamma_H + 2\int_{\mu_{rx}}^m \frac{d\mu}{\mu} \Gamma_J \theta(m-\mu_{rx}) \label{eq:acop} 
 \\ 	& = \frac{\alpha_e}{2\pi} \left[\left( \ln^2 \frac{Q^2}{\mu_{rx}^2} - 3 \ln \frac{Q^2}{\mu_{rx}^2} \right) - \left( \ln^2 \frac{m^2}{\mu_{rx}^2} - \ln \frac{m^2}{\mu_{rx}^2}\right) \theta(m-\mu_{rx})\right].
\end{align}

\begin{figure}[htpb]\centering
    \includegraphics[scale=0.45]{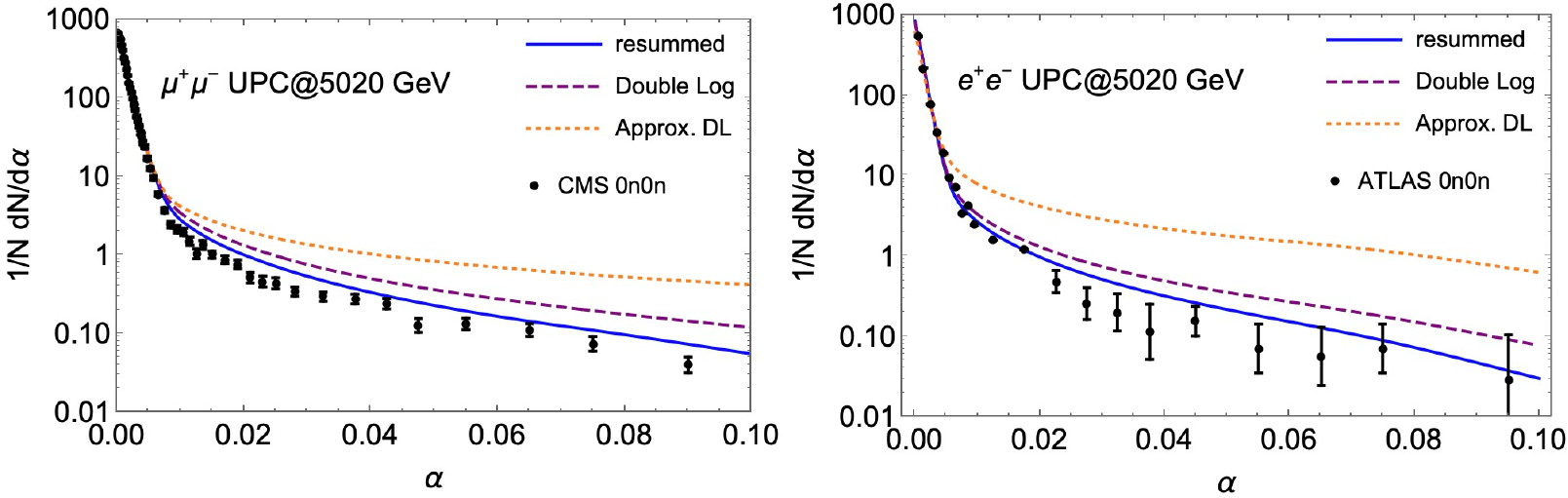}
    \caption{The normalized cross sections of di-lepton production as a function of $\alpha$. Left panel: di-muon production at CMS for the 0n0n case, with the kinematic cutoff: $|y_{1,2}|<2.4$, $P_\perp >3.5$ GeV and $8~\text{GeV}< Q < 60~\text{GeV}$. Right panel: di-electron production at ATLAS for the 0n0n case, with the kinematic cutoff: $|y_{1,2}|<0.8$ and $10~\text{GeV}< Q < 20~\text{GeV}$. The blue solid lines stand for the fully resummed results from Eq.\eqref{eq:acop}, and the purple dashed lines represent the leading double logarithm resummed results obtained using Eq.\eqref{eq:suda_DL}. The acoplanarity distribution reconstructed from the resummed $q_\perp$ distribution given by Eq.~\eqref{eq:res1} and  Eq.~\eqref{eq:sud_DL} is shown with the dotted orange lines. The CMS data displayed in the figure is taken from Ref. \protect\citelow{CMS:2020skx} and the ATLAS data is taken from Ref. \protect\citelow{ATLAS:2022srr}.}
    \label{fig:acoLHC}
\end{figure}

 We numerically calculated the acoplanarity distributions for di-muon production at CMS and di-electron production at ATLAS to compare with their measurements \cite{CMS:2020skx,ATLAS:2022srr}, as shown in Fig.\ref{fig:acoLHC}. We only made estimations for the 0n0n events in which no neutron emission from the excited nuclei in the forward region to avoid possible contributions from incoherent photons. In this case the impact parameter dependence of the cross section is weighted with a $b_\perp$ distribution~(see the review article~\cite{Miller:2007ri} and references therein),
\begin{eqnarray}
2 \pi \int_{2R}^{\infty} b_\perp db_\perp P^2(b_\perp) d \sigma(b_\perp, \ ...),
\end{eqnarray}
 where the probability $P(b_\perp)$ for the 0n event for Pb can be parameterized as~\cite{Baltz:1997di},
\begin{eqnarray}
P(b_\perp)= \exp \left [- \frac{(17.4  \ \text{fm})^2 }{  b_\perp^2 } \right ].
\end{eqnarray}
 
The theoretical results align with both ATLAS and CMS low $\alpha$ data. Yet, in the region of relatively high $\alpha$, our numerical estimations noticeably exceed the experimental data. Compared with the results with double logarithm resummation, the inclusion of the leading single logarithm contribution into the resummation formalism somewhat reduces the discrepancy between the experimental data and the theoretical estimation, although the deviations are still apparent. The potential reason for this needs further investigation. We also reconstruct the  acoplanarity using the resummed $q_\perp$ distribution as given in Eq.~\eqref{eq:res1}. The observed $\alpha$ distribution evidently does not support this approach as shown with the yellow dotted lines in Fig.~\ref{fig:acoLHC}.

\section{Summary}
In heavy ion ultra-peripheral collisions, the photons induced by the relativistically moving ions carry small transverse momenta. In TMD physics, it's known that gluons carrying small transverse momenta in an unpolarized hadron are linearly polarized. Through the analogy of QED to QCD, we derived that coherent photons are highly linearly polarized, and further predicted that this effect can be measured through the $\cos 2\phi$ and $\cos 4\phi$ azimuthal asymmetries in di-lepton production in UPCs. We calculated the azimuthal asymmetries in the RHIC energy region, and consistent data for the $\cos4\phi$ asymmetry in di-electron production process was soon obtained by the STAR collaboration, confirming that coherent photons are linearly polarized.

During the calculation, the contribution from high-order QED corrections cannot be ignored, especially when $q_\perp$ is relatively large. These contributions not only change the size of the cross sections so that change the magnitude of the azimuthal asymmetries, but also will induce azimuthal asymmetries by themselves due to the recoil effect. We revisited the di-lepton production process in UPCs, focusing on the final state soft photon radiation resummation effect on two observables: azimuthal asymmetry and acoplanarity. The results show that at relatively high $q_{\perp}$, the soft photon radiation effect is dominant, and will significantly change the shapes of azimuthal asymmetries, particularly, it will induce large $\cos2\phi$ asymmetry which is absent at leading order due to the negligible electron mass. It is worth mentioning that the resummation of soft photon radiation has different formulas for the $q_\perp$ distribution and acoplanarity, because Fourier transforms were performed in different dimensions. Within the SCET framework, we carry out the resummation for these two observables to all orders, including the contribution of single logarithm. Our results show that the $q_{\perp}$-dependent azimuthal asymmetry is not very sensitive to the sub-leading resummation effect, while the all order one-dimensional resummation is necessary to describe the acoplanarity data of ATLAS and CMS. Although our calculation still has some deviation from the experimental data, the result after including the single logarithm is much closer to the experiment. Therefore, the di-lepton production process in UPCs provides a good opportunity to test the resummation formula through the measurement of angular correlation.

\section*{Acknowledgments}

The author thanks D.Y. Shao, C. Li, C. Zhang and J. Zhou for their contributions to the papers \cite{Li:2019yzy,Li:2019sin,Shao:2023zge} which make this presentation possible.


\section*{References}

\begin{thebibliography}{99}

\bibitem{Mulders:2000sh}
P.~J.~Mulders and J.~Rodrigues,
Phys. Rev. D \textbf{63}, 094021 (2001)
doi:10.1103/PhysRevD.63.094021
[arXiv:hep-ph/0009343 [hep-ph]].

\bibitem{Metz:2011wb}
A.~Metz and J.~Zhou,
Phys. Rev. D \textbf{84}, 051503 (2011)
doi:10.1103/PhysRevD.84.051503
[arXiv:1105.1991 [hep-ph]].

\bibitem{Fermi:1924tc}
E.~Fermi,
Z. Phys. \textbf{29}, 315-327 (1924)
doi:10.1007/BF03184853

\bibitem{vonWeizsacker:1934nji}
C.~F.~von Weizsacker,
Z. Phys. \textbf{88}, 612-625 (1934)
doi:10.1007/BF01333110

\bibitem{Williams:1934ad}
E.~J.~Williams,
Phys. Rev. \textbf{45}, 729-730 (1934)
doi:10.1103/PhysRev.45.729

\bibitem{Li:2019yzy}
C.~Li, J.~Zhou and Y.~J.~Zhou,
Phys. Lett. B \textbf{795}, 576-580 (2019)
doi:10.1016/j.physletb.2019.07.005
[arXiv:1903.10084 [hep-ph]].

\bibitem{Li:2019sin}
C.~Li, J.~Zhou and Y.~J.~Zhou,
Phys. Rev. D \textbf{101}, no.3, 034015 (2020)
doi:10.1103/PhysRevD.101.034015
[arXiv:1911.00237 [hep-ph]].

\bibitem{Shao:2022stc}
D.~Y.~Shao, C.~Zhang, J.~Zhou and Y.~J.~Zhou,
Phys. Rev. D \textbf{107}, no.3, 036020 (2023)
doi:10.1103/PhysRevD.107.036020
[arXiv:2212.05775 [hep-ph]].

\bibitem{Shao:2023zge}
D.~Y.~Shao, C.~Zhang, J.~Zhou and Y.~j.~Zhou,
Phys. Rev. D \textbf{108}, no.11, 116015 (2023)
doi:10.1103/PhysRevD.108.116015
[arXiv:2306.02337 [hep-ph]].

\bibitem{Vidovic:1992ik}
M.~Vidovic, M.~Greiner, C.~Best and G.~Soff,
Phys. Rev. C \textbf{47}, 2308-2319 (1993)
doi:10.1103/PhysRevC.47.2308

\bibitem{Collins:1984kg}
J.~C.~Collins, D.~E.~Soper and G.~F.~Sterman,
Nucl. Phys. B \textbf{250}, 199-224 (1985)
doi:10.1016/0550-3213(85)90479-1

\bibitem{Klein:2018fmp}
S.~Klein, A.~H.~Mueller, B.~W.~Xiao and F.~Yuan,
Phys. Rev. Lett. \textbf{122}, no.13, 132301 (2019)
doi:10.1103/PhysRevLett.122.132301
[arXiv:1811.05519 [hep-ph]].

\bibitem{Hatta:2021jcd}
Y.~Hatta, B.~W.~Xiao, F.~Yuan and J.~Zhou,
Phys. Rev. D \textbf{104}, no.5, 054037 (2021)
doi:10.1103/PhysRevD.104.054037
[arXiv:2106.05307 [hep-ph]].

\bibitem{Adam:2019mby}
J.~Adam \textit{et al.} [STAR],
Phys. Rev. Lett. \textbf{127}, no.5, 052302 (2021)
doi:10.1103/PhysRevLett.127.052302
[arXiv:1910.12400 [nucl-ex]].

\bibitem{Klusek-Gawenda:2020eja}
M.~K\l{}usek-Gawenda, W.~Sch\"afer and A.~Szczurek,
Phys. Lett. B \textbf{814}, 136114 (2021)
doi:10.1016/j.physletb.2021.136114
[arXiv:2012.11973 [hep-ph]].

\bibitem{Klein:2020jom}
S.~Klein, A.~H.~Mueller, B.~W.~Xiao and F.~Yuan,
Phys. Rev. D \textbf{102}, no.9, 094013 (2020)
doi:10.1103/PhysRevD.102.094013
[arXiv:2003.02947 [hep-ph]].

\bibitem{Hatta:2020bgy}
Y.~Hatta, B.~W.~Xiao, F.~Yuan and J.~Zhou,
Phys. Rev. Lett. \textbf{126}, no.14, 142001 (2021)
doi:10.1103/PhysRevLett.126.142001
[arXiv:2010.10774 [hep-ph]].

\bibitem{Catani:2014qha}
S.~Catani, M.~Grazzini and A.~Torre,
Nucl. Phys. B \textbf{890}, 518-538 (2014)
doi:10.1016/j.nuclphysb.2014.11.019
[arXiv:1408.4564 [hep-ph]].

\bibitem{Catani:2017tuc}
S.~Catani, M.~Grazzini and H.~Sargsyan,
JHEP \textbf{06}, 017 (2017)
doi:10.1007/JHEP06(2017)017
[arXiv:1703.08468 [hep-ph]].

\bibitem{Bauer:2000yr}
C.~W.~Bauer, S.~Fleming, D.~Pirjol and I.~W.~Stewart,
Phys. Rev. D \textbf{63}, 114020 (2001)
doi:10.1103/PhysRevD.63.114020
[arXiv:hep-ph/0011336 [hep-ph]].

\bibitem{Bauer:2001ct}
C.~W.~Bauer and I.~W.~Stewart,
Phys. Lett. B \textbf{516}, 134-142 (2001)
doi:10.1016/S0370-2693(01)00902-9
[arXiv:hep-ph/0107001 [hep-ph]].

\bibitem{Bauer:2001yt}
C.~W.~Bauer, D.~Pirjol and I.~W.~Stewart,
Phys. Rev. D \textbf{65}, 054022 (2002)
doi:10.1103/PhysRevD.65.054022
[arXiv:hep-ph/0109045 [hep-ph]].

\bibitem{Bauer:2002nz}
C.~W.~Bauer, S.~Fleming, D.~Pirjol, I.~Z.~Rothstein and I.~W.~Stewart,
Phys. Rev. D \textbf{66}, 014017 (2002)
doi:10.1103/PhysRevD.66.014017
[arXiv:hep-ph/0202088 [hep-ph]].

\bibitem{Beneke:2002ph}
M.~Beneke, A.~P.~Chapovsky, M.~Diehl and T.~Feldmann,
Nucl. Phys. B \textbf{643}, 431-476 (2002)
doi:10.1016/S0550-3213(02)00687-9
[arXiv:hep-ph/0206152 [hep-ph]].

\bibitem{CMS:2020skx}
A.~M.~Sirunyan \textit{et al.} [CMS],
Phys. Rev. Lett. \textbf{127}, no.12, 122001 (2021)
doi:10.1103/PhysRevLett.127.122001
[arXiv:2011.05239 [hep-ex]].

\bibitem{ATLAS:2022srr}
G.~Aad \textit{et al.} [ATLAS],
JHEP \textbf{2306}, 182 (2023)
doi:10.1007/JHEP06(2023)182
[arXiv:2207.12781 [nucl-ex]].

\bibitem{Miller:2007ri}
M.~L.~Miller, K.~Reygers, S.~J.~Sanders and P.~Steinberg,
Ann. Rev. Nucl. Part. Sci. \textbf{57}, 205-243 (2007)
doi:10.1146/annurev.nucl.57.090506.123020
[arXiv:nucl-ex/0701025 [nucl-ex]].

\bibitem{Baltz:1997di}
A.~J.~Baltz and M.~Strikman,
Phys. Rev. D \textbf{57}, 548-549 (1998)
doi:10.1103/PhysRevD.57.548
[arXiv:hep-ph/9705220 [hep-ph]].
\end{thebibliography}

\end{document}